\documentclass[runningheads]{llncs}
\usepackage[T1]{fontenc}
\usepackage{longtable}
\usepackage{graphicx}
\usepackage{arydshln}
\usepackage{hyperref}
\usepackage{url}

\begin{document}
\title{Automated 3D Tumor Segmentation using Temporal Cubic PatchGAN (TCuP-GAN)}
\titlerunning{TCuP-GAN for 3D Brain Tumor Segmentation}
%
\author{Kameswara Bharadwaj Mantha\inst{1} \and
Ramanakumar Sankar\inst{1}\and
Lucy Fortson\inst{1}}
\authorrunning{K.B. Mantha et al.}
%
\institute{Department of Physics \& Astronomy, University of Minnesota Twin Cities, Minneapolis, MN, 55455, USA.\\
\email{manth145@umn.edu}}
\maketitle              
\begin{abstract}
Development of robust general purpose 3D segmentation frameworks using the latest deep learning techniques is one of the active topics in various bio-medical domains. In this work, we introduce Temporal Cubic PatchGAN (TCuP-GAN), a volume-to-volume translational model that marries the concepts of 
a generative feature learning framework with 
Convolutional Long Short-Term Memory Networks (LSTMs), for the task of 3D segmentation. We demonstrate the capabilities of our TCuP-GAN on the data from four segmentation challenges (Adult Glioma, Meningioma, Pediatric Tumors, and Sub-Saharan Africa subset) featured within the 2023 Brain Tumor Segmentation (BraTS) Challenge and quantify its performance using LesionWise Dice similarity and $95\%$ Hausdorff Distance metrics. We demonstrate the successful learning of our framework to predict robust multi-class segmentation masks across all the challenges. This benchmarking work serves as a stepping stone for future efforts towards applying TCuP-GAN on other multi-class tasks such as multi-organelle segmentation in electron microscopy imaging.

\end{abstract}

\section{Introduction \& Motivation}
Brain tumors such as Gliomas and Meningiomas make up some of the most common and aggressive forms of cancers in both adult and pediatric populations, and are quite challenging to treat. Diagnosis and management of these tumors intimately depends on imaging guided techniques, notably using Magnetic Resonance Imaging (MRI) to identify and delineate the extent of the different tumor components. As such, study of these tumors along with development of automated methods for tumor segmentation has been 
a major epicenter for active research. Simultaneously, automated identification and segmentation of different types of cellular organelles within microscopy imaging of different organ tissues is also an open challenge in the domain of bio-medical research and clinical diagnostic applications.

Efficiently acquiring annotated labels for such 3D volume data sets and building considerable sample sizes is one of the critical steps towards achieving generalized and robust models. Citizen Science (CS) techniques implemented through platforms such as the Zooniverse  \cite{Trouille2019}
have become an established method in tackling these issues \cite{Trouille2019}, demonstrating incredible success in various domains and task types including annotating imaging data for bio-medical research purposes \cite{Spiers2021}. 
However, given the ever increasing volume of biomedical datasets (such as from volume electron micrographs or MRI scans), there is a need to accelerate the label gathering step,
while still retaining accuracy.
One of the promising pathways towards efficient and accelerated procurement of annotations and development of generalized and robust machine models is to use ``human-in-the-loop'' strategies where the humans provide corrections to the machine proposal. As such, development and testing of general-purpose models using novel deep learning concepts and making them available for use on future data sets as a starting point for citizen science efforts is a critical necessity, especially in the paradigm of big data.

Through this work, we aim to leverage the different individual deep learning concepts such as 3D Convolutional Neural Networks (CNNs), U-Nets (e.g., \cite{ronneberger2015u,falk2019u}), Generative Adversarial Networks (GANs; \cite{goodfellow14,isola2017image}), and Convolutional Long Short-Term Memory Networks (ConvLSTMs; \cite{shi2015convolutional}) towards designing a general purpose 3D volume-to-volume translation framework that can learn robust generalized spatial features and sequential correlations, and make segmentation predictions. Our inspiration stems from drawing parallels between the fact that MRI images are 3-dimensional and its third spatial axis can be considered analogous to temporal dimension of video datasets. As such, we anticipate that using ConvLSTMs can be helpful in learning 3D features corresponding to the targets of interest (e.g., lesions/tumors).

\section{Related Works}
In the past decade, advancements in deep learning and specifically CNNs \cite{lecun2015deep} have enabled new avenues for automated image segmentation. Several frameworks such as U-Nets \cite{ronneberger2015u} and mask-RCNNs \cite{he2017} have been used for 2D segmentation tasks on not only the general terrestrial datasets \cite{puri19,freudenberg2019large,laxman2021efficient}, but also biomedical applications such as the segmentation of nuclei \cite{vuola2019mask}, blood vessels \cite{xiancheng2018retina}, lung lesions \cite{shaziya2018automatic}, and brain tumors \cite{dong2017automatic}. 

The latest techniques such as 
GANs have been demonstrated as robust image-level feature learners in computer vision tasks \cite{goodfellow14,goodfellow20}. Recently, both 2D and 3D CNNs within GAN methodologies have been paired with U-Net \cite{ronneberger2015u} style architectures to be successful at several computer vision  (e.g., PatchGAN; see \cite{isola2017image}) and biomedical  segmentation tasks (e.g., \cite{dong2019,nema2020rescuenet,choi2019self}). Simultaneously, 
LSTMs \cite{hochreiter1997long} have been demonstrated to be robust learners of sequentially relationed data in text-based corpora, and these concepts  have been successfully extended to the image domain with 2D Convolutional LSTMs (2DConvLSTMs; \cite{shi2015convolutional}) that can learn spatio-sequential features and respective correlations. 

Recently, 2DConvLSTMs have been successfully applied to general segmentation tasks in computer vision, e.g., video semantic segmentation \cite{zhou2021convlstm,qiu2017learning}. The concepts of 2DConvLSTMs have been successfully paired with the U-Net style architectures for the purpose of image segmentation, with extensive exploration in various bio-medical domain related problems (liver lesions \cite{li2021study}, retinal vasculature \cite{yi2022bcr}, and cells within microscopy imaging \cite{arbelle2022dual}). Furthermore, 2DConvLSTMs have also been paired with the U-Net and GAN concepts (e.g., BLU-GAN; \cite{lin2021blu}) for biomedical imaging segmentation. 

Various architectures used in these studies highlight the different methodological choices by which the concepts of 2DConvLSTMs, GANs, and U-Net frameworks have been married. For example, \cite{arbelle2022dual} uses 2DConvLSTM layers within the encoder stage of the U-Net architecture, whereas \cite{yi2022bcr,rani2021disease,lin2021blu} apply them during the encoder-decoder skip connection stage of the U-Net. Similarly, \cite{lin2021blu} incorporated adversarial training concepts with a traditional convolutional discriminator in their 2DConvLSTM-GAN framework, and studies such as \cite{yang2022generative,abramian2019generating} used 3D CNN based PatchGANs on biomedical domain tasks. As such, in this work we are motivated by the unexplored avenue (to the best of our knowledge) of using 2DConvLSTMs in both the encoder and decoder stages of the U-Net along with a 3D PatchGAN framework for biomedical segmentation tasks.

\section{Data}
\label{sec:data}
In this work, we demonstrate the development and application of the TCuP-GAN on four BraTS 2023 segmentation challenge  datasets. Specifically, we use the Adult Glioma (GLI; \cite{baid2021rsna,menze2014multimodal,lloyd2017high}), Meningioma (MEN; \cite{labella2023asnrmiccai}), Pediatric Tumor (PED; \cite{kazerooni2023brain}), and Sub-Saharan Africa (SSA;\cite{adewole2023brain}) training and validation data made accessible via the Synapse platform. The GLI dataset comes with the largest training (validation) set amongst all with $1295$ ($219$) MRI scans, followed by MEN with $1000$ ($141$) scans, and PED, and SSA  with $100$ ($45$), and $60$ ($15$) scans, respectively. Additionally, for ``internal validation'' of our model training, we use a train vs. internal-validation split of $90\%:10\%$ across all the challenge data sets. 


Each data sample across the different challenges is provided in a NIfTI (.nii) format with imaging across four MRI sequences: T1-weighted, T2-weigthed, T1-Contrast Enhanced (T1CE), and Flair with dimensions ($240\times 240\times 155$). Additionally, each sample's annotated segmentation map is also provided, where the Necrotic Tumor (NC), Edematous Region (ED), and Enhanced Tumor (ET) are labelled as classes $1$, $2$, and $3$, respectively. 

For our model training, we perform on-the-fly data pre-processing for each data sample where we normalize each channel data by its 99th percentile pixel value and concatenate the classes together to create a 4 channel image cube. We then resize them to a size of $256\times256$ on their $x,y$ axes, making each sample a 4D image cube with dimensions $(155 \times 4\times 256 \times 256)$. Additionally, to introduce stochastic variation in the input and enable robust feature learning, we also add a random Gaussian noise to each input cube with $\mu = 0$ and $\sigma = 0.1$ to only the pixels that do not correspond to the background (i.e., pixels with values always $>0$).  We perform one-hot encoding of the segmentation map with three classes ($1,2,3$) corresponding to the aforementioned NC, ED, and ET classes. Each of the encoded class masks has pixel values $1$ for regions corresponding to that particular class. After applying the same resizing method, our target segmentation map has the dimensionality of $(155 \times 3\times 256 \times 256)$.

        


\section{Method}

\subsection{Architecture}
Our Temporal Cubic Patch-GAN (TCuP-GAN) is inspired by a generative Image-to-Image translational model (PatchGAN; \cite{isola2017image}), which was introduced for performing 2D semantic segmentation of terrestrial datasets. The PatchGAN framework comprises a 2D U-Net generator that predicts an output segmentation mask for a corresponding input image, along with a 2D convolutional patch-wise discriminator that predicts a 2D discriminatory score map for a concatenated image-mask pair input. Additionally, a traditional U-Net framework  consists of an encoder portion, which extracts features from the input data and compresses the learnt feature space, and a decoder which uses the compressed representation to generate an output target image. Skip connections are used to transfer the learnt features from the encoder to the decoder, thereby making it easier for the model to learn features specific to the output classes, and stabilize the loss landscape \cite{Wang2020}.

\begin{figure}[h!]
    \centering
    \includegraphics[width=0.95\columnwidth]{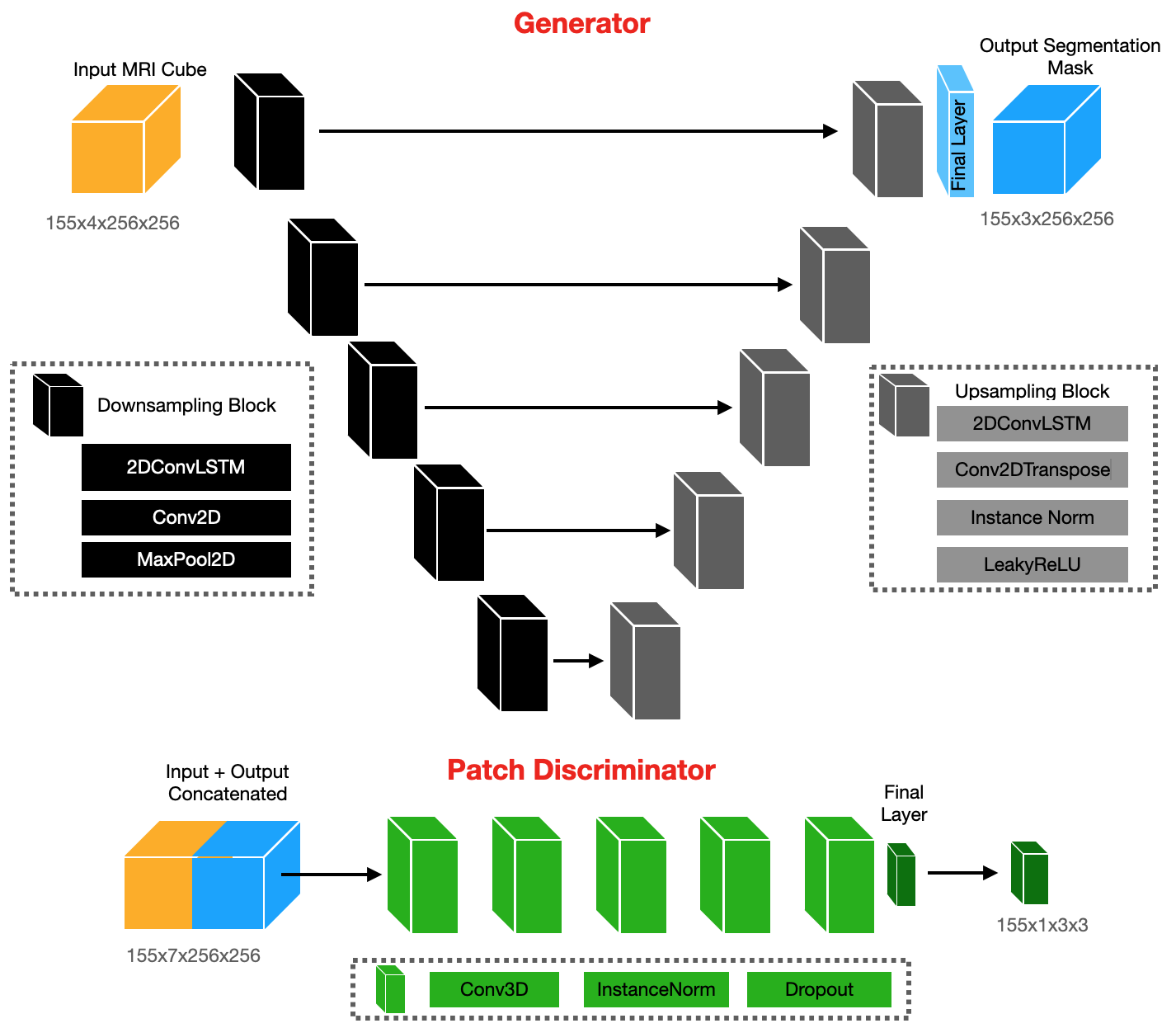}
    \caption{Architecture of our TCuP-GAN framework with its generative component (top) and discriminator (bottom). Different components of our framework are shown in the legend.}
    \label{fig:generator_architecture}
\end{figure}

Our model implementation comprises two components (see Figure\,\ref{fig:generator_architecture}) -- 1) A U-Net based Generator that accepts an image cube input and produces a corresponding $k$-class mask for each depth slice in the input image; and 2) A 3D Convolutional patch-wise Discriminator that takes a concatenation of the input image cube and its corresponding ground truth or generated mask and outputs a $3\times3$ probability matrix per depth slice. 

\noindent{\bf Generator:} Our Generator features an extension over the traditional U-Net generator by replacing the 2D Convolutional layers with 2DConvLSTM \cite{shi2015convolutional} layers, which are an adaption of the traditional fully connected LSTM layers by making use of 2D Convolution blocks instead of the regular multi-layer perceptrons. These are vital in simultaneously capturing the 2D spatial features and the correlation of these features across the depth axis.
The feature vectors ($h_t$) and the cell state ($c$) learned by the 2DConvLSTM encodes the spatial features as a function of 
the third dimension (depth axis for image cubes, or temporal axis for videos),
and the cumulative spatial locations of 
third dimension correlations of these features, respectively. 


Following the input, the encoder portion of the generator is designed to extract and bottleneck $h_t$ while preserving the depth axis using $5$ down-sampling blocks with incremental number of filters (16, 32, 48, 64, 128). Each block consists of a 2DConvLSTM layer (kernel size $3\times 3$) that outputs $h_t$ and $c$, followed by a downsampling unit that contains a 2D Convolutional layer (kernel size $=3\times 3$) along with a MaxPooling layer (pool size $=2\times 2$) that act on spatial dimensions of $h_{t}$. As such, with an input dimensionality (4, 155, 256, 256), the output of the final block is of dimensions (128, 155, 8, 8).

In the decoder portion of the generator, we expand on the output downsampled $h_{t}$ feature vector of the encoder with a series of $5$ upsampling blocks. Each of these blocks contains a 2DConvLSTM layer (kernel size $3\times 3$), where we skip both the $c$ and $h_t$ vectors across the bottleneck from the encoder side. We concatenate this skipped $h_t$ with the output $h_t$ from the previous decoder block before passing it into the next. In each upsampling block, both the $h_{t}$ and the spatial dimensions of the feature vector are upsampled using ConvTranspose layers, followed by Instance Normalization, and LeakyReLU activation. Finally, after the $5$ upsampling decoder blocks, we have an additional Convolutional layer (kernel size $3\times 3$; stride $1\times 1$) that acts on the spatial dimensions, followed by a final Sigmoid activation layer. Our full generator has $2.3$M parameters and its architecture is shown in Figure~\ref{fig:generator_architecture}. 

\noindent{\bf Discriminator}: Our discriminator is a 3D CNN patch-wise binary classifier. Following the input layer, it has a series of $4$ downsampling blocks with the following filter sizes (16, 32, 64, 128), where each block comprises a 3D Convolutional layer (kernel size $1\times 3\times3$), followed by Instance Normalization,  Tanh activation, and Dropout ($=0.2$) layers. Our fifth and final block performs a 3D Convolution (kernel size $1\times 3\times3$) and  has a Sigmoid activated single channel output. Each unit of the output of the discriminator represents a concatenation of 3D patch of the input and output image cubes, and provides the probability that the image + mask patch is real.  Figure~\ref{fig:generator_architecture} shows the architecture of the discriminator.

\subsection{Training Strategy}
In this study, we develop and apply our TCuP-GAN framework on four BraTS 2023 challenge data sets. Here, we describe our general training strategy and hyper-parameters used. Furthermore, as mentioned earlier, some of the challenges come with a relatively small number of training samples (see \S\,\ref{sec:data}). As such, we chose to use transfer learning of the model trained on the largest sample size data (GLI) to the others (MEN, PED, SSA).

We train our TCuP-GAN model from scratch on the GLI data set for $30$ epochs (with a batch size of $2$) by minimizing a scaled, class-weighted Binary Cross-Entropy loss function $BCE{^\prime} = \gamma \times \mathcal{W} \times BCE$ between the ground truth (GT) and predicted (PD) segmentation mask, where $\mathcal{W} = 1 - \sum_{d,x,y}(GT)/\sum(GT)$. We use $\gamma = 200$ to stabilize the network training at low $BCE$ value regimes. We employ a starting learning rate of $r = 5\times 10^{-4}$ and $r = 10^{-4}$ for our generator and discriminator, respectively, and decay the learning rates as $r^{0.95}$ every $5$ epochs. We find that the GLI model achieved a stable loss at the end of its training with near identical mean $BCE^{\prime}$ on the internal validation samples. As for the training on the MEN, PED, and SSA challenge data, we initialize the model with GLI-based model weights (at 30th epoch) and further train it for $30$ more epochs with a smaller (constant) learning rate of $r = 10^{-4}$ for the generator (and $r = 10^{-4}$ for the discriminator).

\begin{figure}[h!]
    \centering
    \includegraphics[width=\columnwidth]{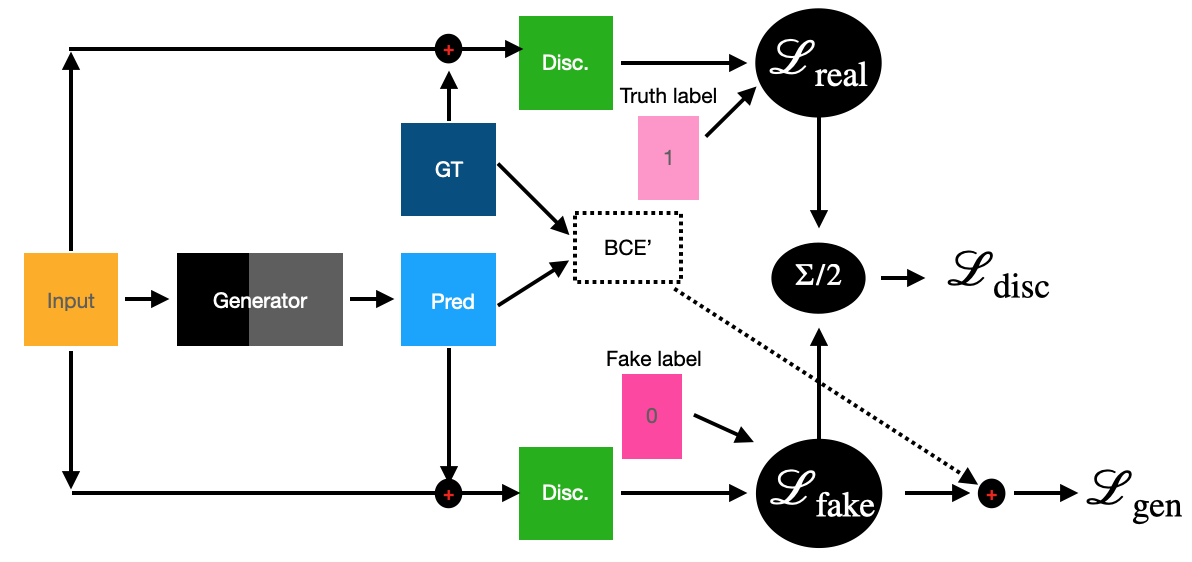}
    \caption{Illustration of the optimization strategy of our overall generator and discriminator frameworks.}
    \label{fig:strategy}
\end{figure}

In each training step (i.e., for each batch), the input cube is passed through the generator to yield the predicted mask and the $BCE^{\prime}$ loss is computed between them. Additionally, the concatenated pairs (input, GT) and (input, PD) are passed through the discriminator and two (BCE) losses $\mathcal{L}_{\rm real}$ and $\mathcal{L_{\rm fake}}$ are computed between the discriminator outputs and corresponding real (unity), fake (zero) matrices, respectively. The total discriminator loss is defined as the average of these aforementioned losses ($\mathcal{L}_{\rm disc} = [\mathcal{L}_{\rm real} + \mathcal{L}_{\rm fake}]/2$). The total generator loss is defined as $\mathcal{L}_{\rm gen} = BCE^{\prime} + \mathcal{L}_{\rm fake}$.  In Figure\,\ref{fig:strategy}, we visually show the generator and discriminator loss computation strategy.


\section{Evaluation}
Thus far, we described the design and training optimization strategy of the TCuP-GAN model on data from four datasets featured as a part of the BraTS 2023 challenge. In this section, we describe the post-processing of our model's output 3D segmentation masks and quantification of the standard metrics that describe the model performance.  

\subsection{Model Performance Assessment Metrics}
Following the recommendations of the BraTS 2023 challenge evaluation procedures, we quantify the performance of our models using the following core metrics: `LesionWise' Dice ($\mathcal{D}_{\rm lw}$) and Hausdorff distances ($\mathcal{H}_{95,lw}$) \cite{bakas2017advancing,bakas2017segmentation}, where the average Dice and Hausdorff distance metrics are computed across individual, connected volume regions identified by binary connectivity (referred to as ``lesion''), while also penalizing false positive lesion occurrences (with $\mathcal{D}_{\rm lw}=0$ and $\mathcal{H}_{95,lw}=374$). We use the official BraTS 2023 evaluation pipeline\footnote{\href{https://github.com/rachitsaluja/brats\_val\_2023/}{\url{https://github.com/rachitsaluja/brats\_val\_2023/}}} to compute these metrics for our internal validation set and present the yielded results after submitting our predicted mask on the provided validation sets. It is worth noting that although the models are trained on the native classes provided by the data sets (NC, ED, and ET),  we report the metrics on the ``Whole Tumor'' (WT $=$ TC $+$ ED), with ``Tumor Core'' (TC $=$ ET $+$ NT), and ``Enhanced Tumor'' (ET).

\subsection{Post-processing}
For each input MRI volume, our trained model outputs a 3D segmentation mask with each voxel denoting the probability of that voxel being a given class (with dimensions $155\times 3 \times 256 \times 256$), where the second dimension corresponds to the 3 different classes as defined in \S\,\ref{sec:data}. Our first step of post-processing involves defining a ``background'' mask as the $1 - \Sigma_{1,2,3} {\rm (PD)}$ and concatenating it with PD, yielding a $4$-channel mask of dimensions ($155\times 4 \times 256 \times 256$). Next, we normalize the PD by dividing it with its sum along the class axis and apply the {\tt argmax()} operation to the model output along it, thereby yielding a ``combined'' segmentation mask where each voxel is labelled one of (0,1,2,3) for background, NC, ED, and ET, respectively. 

Next, we further process each combined mask to reject any spurious small-volume regions. This refinement step involves first splitting the combined mask into the BraTS challenge relevant classes -- WT, TC, and ET. Next, for each of these three classes, we generate a binary connectivity map (using the {\it connected-components-3d} library), where connected voxels are uniquely labeled. We then retain those regions that have a mean area (across depth slices) greater than a threshold value ($\mathcal{A}_{\rm thresh}$) and span at least $5$ depth slices. For each model trained on its corresponding BraTS challenge dataset, we run an iterative Monte-Carlo experiment, where we vary the class-wise $\mathcal{A}_{\rm thresh}$ and chose an optimal value that yielded highest $\mathcal{D}_{\rm lw}$ on the Internal Validation data. For the GLI, MEN, PED, and SSA models, the chosen class-wise $\mathcal{A}_{\rm thresh}$ values (WT, TC, ET) are (125, 75, 20), (125, 125, 25), (75, 75, 25), and (75, 100, 5), respectively.



\section{Results \& Discussion}
\begin{figure}
    \centering
    \includegraphics[width=\columnwidth]{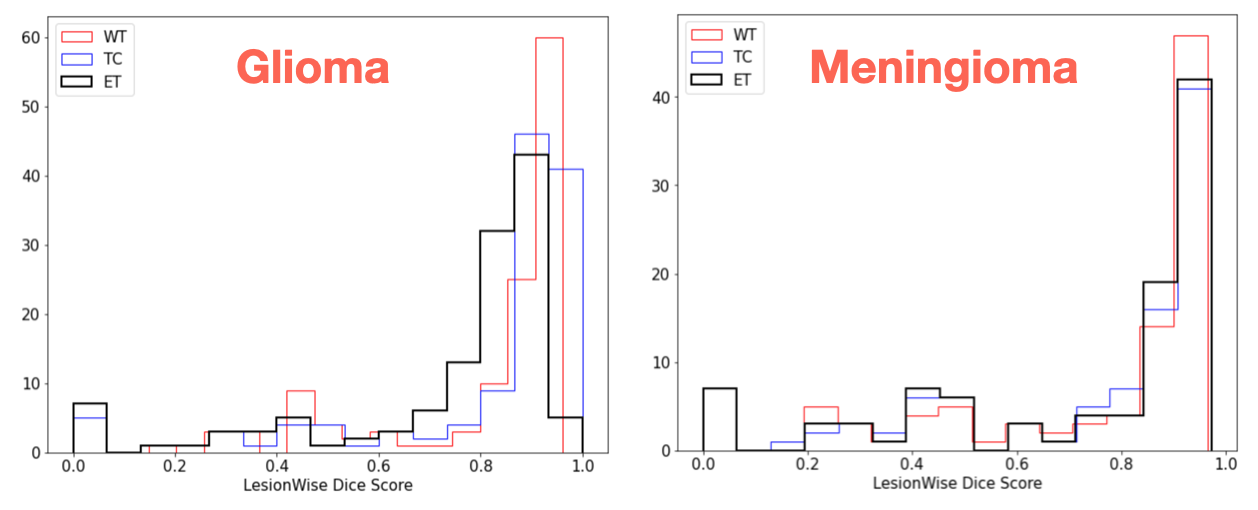}
    \caption{Distributions of the lesionwise dice ($\mathcal{D}_{\rm lw}$ based on the GLI and MEN models applied to the internal validation set.}
    \label{fig:gli_men_histograms}
\end{figure}

In this section, we report and discuss the mean and median lesionwise metrics $\mathcal{D}_{\rm lw}$ and $\mathcal{H}_{\rm 95, lw}$ for the different BraTS challenge datasets (reported in Table\,\ref{tab:summary_stats}). Specifically, we report these metrics for both our internal validation set and the team provided validation set.
\begin{longtable}{ccc|c|c|c}
    \caption{Summary of the LesionWise Dice ($\mathcal{D}_{\rm lw}$) and $95\%$ Hausdorff Distance ($\mathcal{H}_{95\%}$) metrics for the different evaluation datasets across different challenge datasets. The internal validation sample based results are reported in parentheses. }
    \label{tab:summary_stats}\\
    \hline
    Model Name & Eval. Dataset & Statistic Name & Lesion & Mean & Median  \\
    \hline
    &  &  & WT & 0.83 (0.81) & 0.9 (0.9)\\
     &  & $\mathcal{D}_{\rm lw}$ & TC & 0.76 (0.8) & 0.9 (0.91)\\
     & & & ET & 0.76 (0.73) & 0.84 (0.83)\\
    GLI & (Internal) Validation & & & & \\
     & &  & WT & 29.3 (40.9) & 4.1 (4.2)\\
     & & $\mathcal{H}_{\rm 95,lw}$ & TC & 43.5 (34.3) & 3.3 (2.2)\\
     & & & ET & 35.6 (43.3) & 2.0 (1.7)\\
    \hline
    &  &  & WT & 0.67 (0.72) & 0.85 (0.88) \\
     &  & $\mathcal{D}_{\rm lw}$ & TC & 0.68 (0.73)  & 0.85 (0.88) \\
     & & & ET & 0.68 (0.73) & 0.85 (0.88) \\
    MEN & (Internal) Validation & & & & \\
     & &  & WT & 86.8 (66.7) & 3.69 (3.0) \\
     & & $\mathcal{H}_{\rm 95,lw}$ & TC & 82.2 (65.5) & 3.16 (2.2)\\
     & & & ET & 81.8 (65.0) & 2.23 (2.0)\\
     \hline
     &  &  & WT & 0.74 (0.70) & 0.84 (0.87)\\
     &  & $\mathcal{D}_{\rm lw}$ & TC & 0.66 (0.64) &  0.75 (0.76)\\
     & & & ET & 0.45 (0.86) & 0.36 (0.80) \\
    PED & (Internal) Validation & & & & \\
     & &  & WT & 48.8 (77.6) & 6.1 (8.0) \\
     & & $\mathcal{H}_{\rm 95,lw}$ & TC & 58.7 (89.2) & 9.2 (8.0)\\
     & & & ET & 152.9 (1.6) & 14.9 (1.4)\\
     \hline
     &  &  & WT & 0.67 (0.63) & 0.73 (0.75) \\
     &  & $\mathcal{D}_{\rm lw}$ & TC & 0.64 (0.44) & 0.83 (0.48) \\
     & & & ET & 0.63 (0.42) & 0.78 (0.47) \\
    SSA & (Internal) Validation & & & & \\
     & &  & WT & 76 (96.6) & 6.92 (24.5) \\
     & & $\mathcal{H}_{\rm 95,lw}$ & TC & 74.7 (161.3) & 6.1 (102.6)\\
     & & & ET & 72 (159.5) & 4.0 (101.6)\\
    \hline
    
\end{longtable}
Generally speaking, we find that our TCuP-GAN framework successfully learns to predict the target multi-class segmentation labels across different datasets. With the reported metrics based on the validation set in context, we note that all the models demonstrated a good performance on the WT class with the mean $\mathcal{D}_{\rm lw} \sim 0.65-0.8$. As for the TC and ET classes, the GLI model showed the highest performance $\mathcal{D}_{\rm lw} \sim 0.76$, whereas the MEN, PED, and SSA models performed at $\mathcal{D}_{\rm lw} \sim 0.63-0.68$, with the exception of the PED model performing the lowest scores at $\mathcal{D}_{\rm lw} \sim 0.45$ for the ET class. Additionally, across these three classes and the $4$ models, we highlight that the median $\mathcal{D}_{\rm lw}$ values remain substantially higher (ranging between $\sim 0.7-0.9$)  than the reported means. This can be visually comprehended in Figure\,\ref{fig:gli_men_histograms}, where we show the distributions of $\mathcal{D}_{\rm lw}$ and $\mathcal{H}_{\rm 95, lw}$ values for the GLI and  MEN models applied to the Internal Validation sample. We note that while our model tends to performs very well on a dominant portion of the validation set, a few outlier cases where the models yielded a $\mathcal{D}_{\rm lw} \sim 0$ or $\sim 0.5$ impacted the mean $\mathcal{D}_{\rm lw}$ to be smaller. When our model failed to predict any segmentation regions for a few samples (especially for the TC or ET classes), this resulted in a $\mathcal{D}_{\rm lw} \sim 0$. Simultaneously, we noticed that the samples that have $0.1\leq \mathcal{D}_{\rm lw} \leq 0.5$ are because of the model predicting one to two large false-positive (FP) region(s) that incrementally (and asymptotically) down-weight the $\mathcal{D}_{\rm lw}$ (as FP regions are assigned a $\mathcal{D}_{\rm lw} = 0$). For example, in a case where our model predicts a true positive with $\mathcal{D}_{\rm lw} \sim 0.9$ and a false positive, then the averaged $\mathcal{D}_{\rm lw}$ reported is $\sim 0.45$.

The class-wise performance trends of $\mathcal{D}_{\rm lw}$ (and $\mathcal{H}_{\rm 95, lw}$) across different models also mostly follow the same behavior when assessing the internal validation based values, albeit with a worthwhile note that they tend to be sometimes higher when compared to the validation set based results. However, it is to be also noted that the median metric values between the internal validation and validation sets seem to be much more in close agreement with each other than the mean values. This indicates a higher incidence of FPs in the validation set when compared to our internal validation data, however, we are cognizant that such differences are to be expected as our internal validation set is one random realization of the entire sample. Finally, we also highlight that all of our assessments and interpretations based on  $\mathcal{D}_{\rm lw}$ also resonate with  $\mathcal{H}_{\rm 95, lw}$ values in a self-consistent way.


\section{Conclusions}
In this work, we introduced a 3D volume-to-volume translation model -- Temporal Cubic PatchGAN (TCuP-GAN), which  leverages the conceptual frameworks of U-Nets, 2DConvLSTMs, and GANs (especially PatchGAN) for the purpose of automated 3D segmentation of brain tumors. We demonstrated the successful application and performance of our framework on four datasets within the BraTS 2023 segmentation challenge aimed at segmenting Gliomas, Meningiomas, and Pediatric tumors. This benchmarking analysis lays critical groundwork for future applications of TCuP-GAN on different biomedical segmentation tasks (e.g., multi-organelle segmentation in microscopy imaging) and interfacing with citizen science platforms such as Zooniverse as a general-purpose segmentation model for project teams to use in a human-in-the-loop strategy. 

\subsection*{Acknowledgments}
This work was partially supported by the  National Science Foundation under grants OAC 1835530 and IIS 2006894. The authors also acknowledge the Minnesota Supercomputing Institute (MSI) at the University of Minnesota for providing resources that contributed to the research results reported within this paper. URL: http://www.msi.umn.edu.

\newpage
\bibliographystyle{splncs04}
\bibliography{bibliograph}

\begin{thebibliography}{10}
\providecommand{\url}[1]{\texttt{#1}}
\providecommand{\urlprefix}{URL }
\providecommand{\doi}[1]{https://doi.org/#1}

\bibitem{abramian2019generating}
Abramian, D., Eklund, A.: Generating fmri volumes from t1-weighted volumes using 3d cyclegan. arXiv preprint arXiv:1907.08533  (2019)

\bibitem{adewole2023brain}
Adewole, M., Rudie, J.D., Gbadamosi, A., Toyobo, O., Raymond, C., Zhang, D., Omidiji, O., Akinola, R., Suwaid, M.A., Emegoakor, A., et~al.: The brain tumor segmentation (brats) challenge 2023: Glioma segmentation in sub-saharan africa patient population (brats-africa). arXiv preprint arXiv:2305.19369  (2023)

\bibitem{arbelle2022dual}
Arbelle, A., Cohen, S., Raviv, T.R.: Dual-task convlstm-unet for instance segmentation of weakly annotated microscopy videos. IEEE Transactions on Medical Imaging  \textbf{41}(8),  1948--1960 (2022)

\bibitem{baid2021rsna}
Baid, U., Ghodasara, S., Mohan, S., Bilello, M., Calabrese, E., Colak, E., Farahani, K., Kalpathy-Cramer, J., Kitamura, F.C., Pati, S., et~al.: The rsna-asnr-miccai brats 2021 benchmark on brain tumor segmentation and radiogenomic classification. arXiv preprint arXiv:2107.02314  (2021)

\bibitem{bakas2017segmentation}
Bakas, S., Akbari, H., Sotiras, A., Bilello, M., Rozycki, M., Kirby, J., Freymann, J., Farahani, K., Davatzikos, C.: Segmentation labels and radiomic features for the pre-operative scans of the tcga-lgg collection. The cancer imaging archive  \textbf{286} (2017)

\bibitem{bakas2017advancing}
Bakas, S., Akbari, H., Sotiras, A., Bilello, M., Rozycki, M., Kirby, J.S., Freymann, J.B., Farahani, K., Davatzikos, C.: Advancing the cancer genome atlas glioma mri collections with expert segmentation labels and radiomic features. Scientific data  \textbf{4}(1),  1--13 (2017)

\bibitem{choi2019self}
Choi, J., Kim, T., Kim, C.: Self-ensembling with gan-based data augmentation for domain adaptation in semantic segmentation. In: Proceedings of the IEEE/CVF International Conference on Computer Vision. pp. 6830--6840 (2019)

\bibitem{dong2017automatic}
Dong, H., Yang, G., Liu, F., Mo, Y., Guo, Y.: Automatic brain tumor detection and segmentation using u-net based fully convolutional networks. In: Medical Image Understanding and Analysis: 21st Annual Conference, MIUA 2017, Edinburgh, UK, July 11--13, 2017, Proceedings 21. pp. 506--517. Springer (2017)

\bibitem{dong2019}
Dong, X., Lei, Y., Wang, T., Thomas, M., Tang, L., Curran, W.J., Liu, T., Yang, X.: Automatic multiorgan segmentation in thorax ct images using u-net-gan. Medical physics  \textbf{46}(5),  2157--2168 (2019)

\bibitem{falk2019u}
Falk, T., Mai, D., Bensch, R., {\c{C}}i{\c{c}}ek, {\"O}., Abdulkadir, A., Marrakchi, Y., B{\"o}hm, A., Deubner, J., J{\"a}ckel, Z., Seiwald, K., et~al.: U-net: deep learning for cell counting, detection, and morphometry. Nature methods  \textbf{16}(1),  67--70 (2019)

\bibitem{freudenberg2019large}
Freudenberg, M., N{\"o}lke, N., Agostini, A., Urban, K., W{\"o}rg{\"o}tter, F., Kleinn, C.: Large scale palm tree detection in high resolution satellite images using u-net. Remote Sensing  \textbf{11}(3), ~312 (2019)

\bibitem{goodfellow14}
Goodfellow, I., Pouget-Abadie, J., Mirza, M., Xu, B., Warde-Farley, D., Ozair, S., Courville, A., Bengio, Y.: Generative adversarial nets. Advances in neural information processing systems  \textbf{27} (2014)

\bibitem{goodfellow20}
Goodfellow, I., Pouget-Abadie, J., Mirza, M., Xu, B., Warde-Farley, D., Ozair, S., Courville, A., Bengio, Y.: Generative adversarial networks. Communications of the ACM  \textbf{63}(11),  139--144 (2020)

\bibitem{he2017}
He, K., Gkioxari, G., Doll{\'a}r, P., Girshick, R.: Mask r-cnn. In: Proceedings of the IEEE international conference on computer vision. pp. 2961--2969 (2017)

\bibitem{hochreiter1997long}
Hochreiter, S., Schmidhuber, J.: Long short-term memory. Neural computation  \textbf{9}(8),  1735--1780 (1997)

\bibitem{isola2017image}
Isola, P., Zhu, J.Y., Zhou, T., Efros, A.A.: Image-to-image translation with conditional adversarial networks. In: Proceedings of the IEEE conference on computer vision and pattern recognition. pp. 1125--1134 (2017)

\bibitem{kazerooni2023brain}
Kazerooni, A.F., Khalili, N., Liu, X., Haldar, D., Jiang, Z., Anwar, S.M., Albrecht, J., Adewole, M., Anazodo, U., Anderson, H., et~al.: The brain tumor segmentation (brats) challenge 2023: Focus on pediatrics (cbtn-connect-dipgr-asnr-miccai brats-peds). arXiv preprint arXiv:2305.17033  (2023)

\bibitem{labella2023asnrmiccai}
LaBella, D., Adewole, M., Alonso-Basanta, M., Altes, T., Anwar, S.M., Baid, U., Bergquist, T., Bhalerao, R., Chen, S., Chung, V., Conte, G.M., Dako, F., Eddy, J., Ezhov, I., Godfrey, D., Hilal, F., Familiar, A., Farahani, K., Iglesias, J.E., Jiang, Z., Johanson, E., Kazerooni, A.F., Kent, C., Kirkpatrick, J., Kofler, F., Leemput, K.V., Li, H.B., Liu, X., Mahtabfar, A., McBurney-Lin, S., McLean, R., Meier, Z., Moawad, A.W., Mongan, J., Nedelec, P., Pajot, M., Piraud, M., Rashid, A., Reitman, Z., Shinohara, R.T., Velichko, Y., Wang, C., Warman, P., Wiggins, W., Aboian, M., Albrecht, J., Anazodo, U., Bakas, S., Flanders, A., Janas, A., Khanna, G., Linguraru, M.G., Menze, B., Nada, A., Rauschecker, A.M., Rudie, J., Tahon, N.H., Villanueva-Meyer, J., Wiestler, B., Calabrese, E.: The asnr-miccai brain tumor segmentation (brats) challenge 2023: Intracranial meningioma (2023)

\bibitem{laxman2021efficient}
Laxman, K., Dubey, S.R., Kalyan, B., Kojjarapu, S.R.V.: Efficient high-resolution image-to-image translation using multi-scale gradient u-net. In: International Conference on Computer Vision and Image Processing. pp. 33--44. Springer (2021)

\bibitem{lecun2015deep}
LeCun, Y., Bengio, Y., Hinton, G.: Deep learning. nature  \textbf{521}(7553),  436--444 (2015)

\bibitem{li2021study}
Li, J., Ou, X., Shen, N., Sun, J., Ding, J., Zhang, J., Yao, J., Wang, Z.: Study on strategy of ct image sequence segmentation for liver and tumor based on u-net and bi-convlstm. Expert Systems with Applications  \textbf{180},  115008 (2021)

\bibitem{lin2021blu}
Lin, L., Wu, J., Cheng, P., Wang, K., Tang, X.: Blu-gan: Bi-directional convlstm u-net with generative adversarial training for retinal vessel segmentation. In: Intelligent Computing and Block Chain: First BenchCouncil International Federated Conferences, FICC 2020, Qingdao, China, October 30--November 3, 2020, Revised Selected Papers 1. pp. 3--13. Springer (2021)

\bibitem{lloyd2017high}
Lloyd, C.T., Sorichetta, A., Tatem, A.J.: High resolution global gridded data for use in population studies. Scientific data  \textbf{4}(1),  1--17 (2017)

\bibitem{menze2014multimodal}
Menze, B.H., Jakab, A., Bauer, S., Kalpathy-Cramer, J., Farahani, K., Kirby, J., Burren, Y., Porz, N., Slotboom, J., Wiest, R., et~al.: The multimodal brain tumor image segmentation benchmark (brats). IEEE transactions on medical imaging  \textbf{34}(10),  1993--2024 (2014)

\bibitem{nema2020rescuenet}
Nema, S., Dudhane, A., Murala, S., Naidu, S.: Rescuenet: An unpaired gan for brain tumor segmentation. Biomedical Signal Processing and Control  \textbf{55},  101641 (2020)

\bibitem{puri19}
Puri, D.: Coco dataset stuff segmentation challenge. In: 2019 5th international conference on computing, communication, control and automation (ICCUBEA). pp.~1--5. IEEE (2019)

\bibitem{qiu2017learning}
Qiu, Z., Yao, T., Mei, T.: Learning deep spatio-temporal dependence for semantic video segmentation. IEEE Transactions on Multimedia  \textbf{20}(4),  939--949 (2017)

\bibitem{rani2021disease}
Rani, B., Ratna, V.R., Srinivasan, V.P., Thenmalar, S., Kanimozhi, R.: Disease prediction based retinal segmentation using bi-directional convlstmu-net. Journal of Ambient Intelligence and Humanized Computing pp. 1--10 (2021)

\bibitem{ronneberger2015u}
Ronneberger, O., Fischer, P., Brox, T.: U-net: Convolutional networks for biomedical image segmentation. In: Medical Image Computing and Computer-Assisted Intervention--MICCAI 2015: 18th International Conference, Munich, Germany, October 5-9, 2015, Proceedings, Part III 18. pp. 234--241. Springer (2015)

\bibitem{shaziya2018automatic}
Shaziya, H., Shyamala, K., Zaheer, R.: Automatic lung segmentation on thoracic ct scans using u-net convolutional network. In: 2018 International conference on communication and signal processing (ICCSP). pp. 0643--0647. IEEE (2018)

\bibitem{shi2015convolutional}
Shi, X., Chen, Z., Wang, H., Yeung, D.Y., Wong, W.K., Woo, W.c.: Convolutional lstm network: A machine learning approach for precipitation nowcasting. Advances in neural information processing systems  \textbf{28} (2015)

\bibitem{Spiers2021}
Spiers, H., Songhurst, H., Nightingale, L., de~Folter, J., Community, T.Z.V., Hutchings, R., Peddie, C.J., Weston, A., Strange, A., Hindmarsh, S., Lintott, C., Collinson, L.M., Jones, M.L.: Deep learning for automatic segmentation of the nuclear envelope in electron microscopy data, trained with volunteer segmentations. Traffic  \textbf{22}(7),  240--253 (2021). \doi{https://doi.org/10.1111/tra.12789}, \url{https://onlinelibrary.wiley.com/doi/abs/10.1111/tra.12789}

\bibitem{Trouille2019}
Trouille, L., Lintott, C.J., Fortson, L.F.: Citizen science frontiers: Efficiency, engagement, and serendipitous discovery with human{\textendash}machine systems. Proceedings of the National Academy of Sciences  \textbf{116}(6),  1902--1909 (2019). \doi{10.1073/pnas.1807190116}, \url{https://www.pnas.org/content/116/6/1902}

\bibitem{vuola2019mask}
Vuola, A.O., Akram, S.U., Kannala, J.: Mask-rcnn and u-net ensembled for nuclei segmentation. In: 2019 IEEE 16th international symposium on biomedical imaging (ISBI 2019). pp. 208--212. IEEE (2019)

\bibitem{Wang2020}
Wang, L., Shen, B., Zhao, N., Zhang, Z.: Is the skip connection provable to reform the neural network loss landscape? In: IJCAI (2020)

\bibitem{xiancheng2018retina}
Xiancheng, W., Wei, L., Bingyi, M., He, J., Jiang, Z., Xu, W., Ji, Z., Hong, G., Zhaomeng, S.: Retina blood vessel segmentation using a u-net based convolutional neural network. In: Procedia Computer Science: International Conference on Data Science (ICDS 2018). pp.~8--9 (2018)

\bibitem{yang2022generative}
Yang, C.J., Lin, C.L., Wang, C.K., Wang, J.Y., Chen, C.C., Su, F.C., Lee, Y.J., Lui, C.C., Yeh, L.R., Fang, Y.H.D.: Generative adversarial network (gan) for automatic reconstruction of the 3d spine structure by using simulated bi-planar x-ray images. Diagnostics  \textbf{12}(5), ~1121 (2022)

\bibitem{yi2022bcr}
Yi, Y., Guo, C., Hu, Y., Zhou, W., Wang, W.: Bcr-unet: Bi-directional convlstm residual u-net for retinal blood vessel segmentation. Frontiers in Public Health  \textbf{10},  1056226 (2022)

\bibitem{zhou2021convlstm}
Zhou, L., Yuan, H., Ge, C.: Convlstm-based neural network for video semantic segmentation. In: 2021 International Conference on Visual Communications and Image Processing (VCIP). pp.~1--5. IEEE (2021)

\end{thebibliography}

\end{document}